\documentclass[%
 reprint,
 amsmath,amssymb,
 aps,
 prl,
]{revtex4-1}

\usepackage{graphicx}
\usepackage{dcolumn}
\usepackage{bm}
\usepackage{hyperref}
\usepackage{textgreek}

\begin{document}

\preprint{APS/123-QED}


\title{Self-switching Kerr oscillations of counter-propagating light in microresonators}

\author{Michael T. M. \surname{Woodley$^{1,2,\dagger}$}}
\email{michael.woodley@npl.co.uk}
\author{Lewis \surname{Hill$^{1,3,\dagger}$}}
\author{Leonardo \surname{Del Bino$^{1,2}$}}
\author{Gian-Luca \surname{Oppo$^3$}}
\author{Pascal \surname{Del'Haye$^{4,5}$}}
\affiliation{$^1$National Physical Laboratory, Hampton Road, Teddington, TW11 0LW, UK\\$^2$ SUPA and Department of Physics, Heriot-Watt University, Edinburgh, EH14 4AS, UK\\$^3$SUPA and Department of Physics, University of Strathclyde, 107 Rottenrow, Glasgow, G4 0NG, UK\\$^4$Max Planck Institute for the Science of Light, Staudtstr. 2, 91058 Erlangen, Germany\\$^5$Department of Physics, Friedrich Alexander University Erlangen-Nuremberg, 91058 Erlangen, Germany}

\begin{abstract}
We report the experimental observation of oscillatory antiphase switching between counter-propagating light beams in Kerr ring microresonators, including the emergence of periodic behaviour from a chaotic regime. Self-switching occurs in balanced regimes of operation and is well captured by a simple coupled dynamical system featuring only the self- and cross-phase Kerr nonlinearities. Switching phenomena are due to temporal instabilities of symmetry-broken states combined with attractor merging that restores the broken symmetry on average. Self-switching of counter-propagating light is robust for realising controllable, all-optical generation of waveforms, signal encoding and chaotic cryptography.
\end{abstract}

\maketitle



\begin{figure}[htbp]
\centering
\fbox{\includegraphics[width=0.95\linewidth]{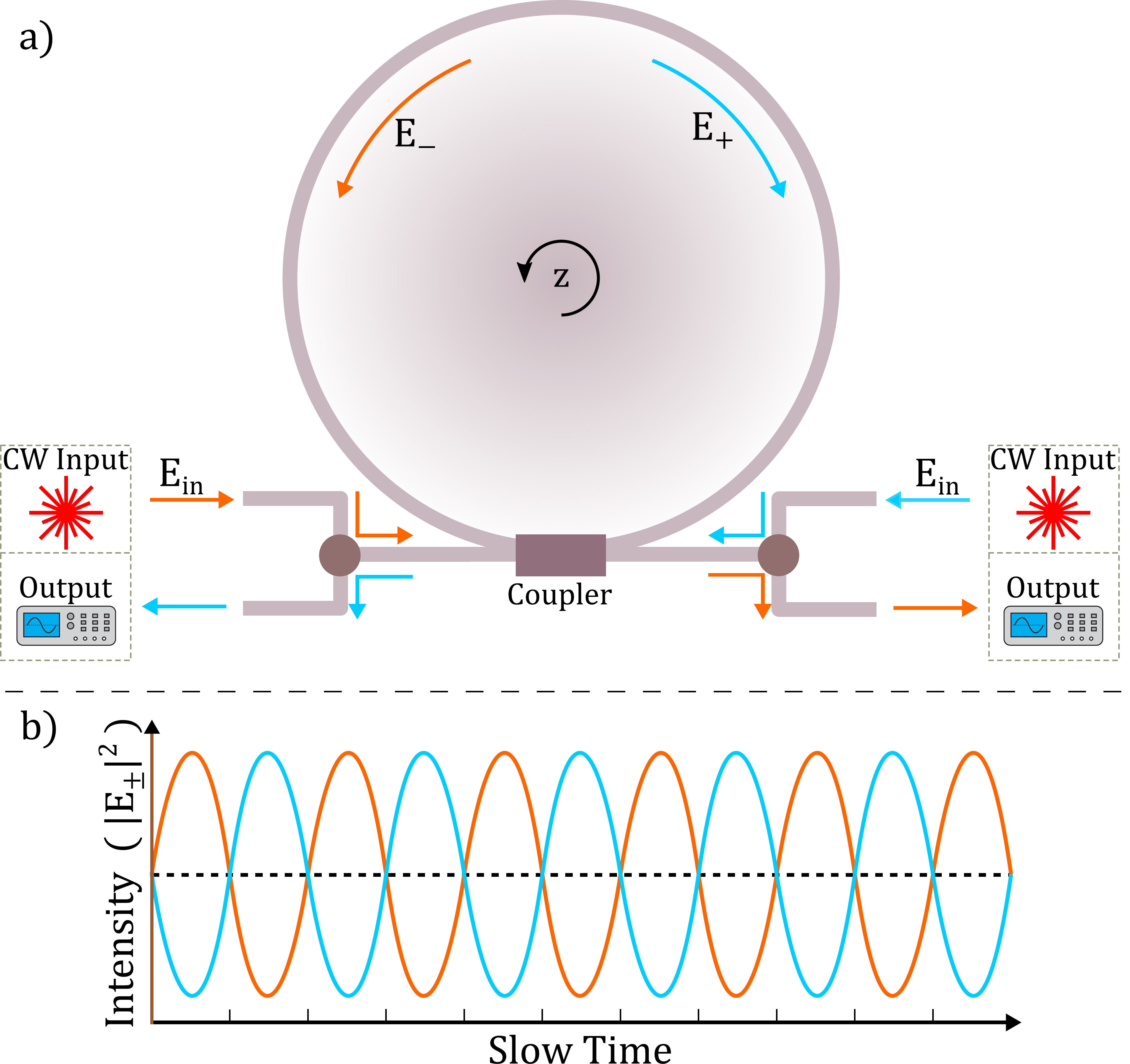}}
\caption{a) Schematic of a ring resonator. Two identical input beams enter the resonator, via a coupling mechanism, to then circulate in opposing directions. They each complete many round-trips before eventually leaving the resonator to continue to their respective outputs. b) Example of antiphase periodic switching. ``Slow time" refers to a time-scale much larger than the round-trip time of the resonator.}
\label{fig:CounterFig}
\end{figure}

In 1981, Kaplan and Meystre theoretically studied Kerr ring resonators pumped with two input optical beams, of equal intensity and frequency, which enter the resonator in opposing directions \cite{kaplan1982directionally} -- see Fig.~\ref{fig:CounterFig}a). Resonant build-up then occurs, and the intracavity fields interact through Kerr cross-phase modulation.

A key phenomenon that can occur in this bidirectionally-pumped system above a certain threshold is a spontaneous symmetry breaking of the circulating field intensities -- i.e. the sudden change from equal intensities to a state with one field is dominant and the other is suppressed. This has been subject to much investigation \cite{kaplan1981enhancement, kaplan1982directionally,del2018microresonator,del2017symmetry,wright1985theory,woodley2018universal,otsuka1983nonlinear,firth1988transverse,Wu2019, Copie2019}. Microresonators can additionally exhibit oscillatory behaviours due to a variety of mechanisms such as thermal instabilities \cite{Fomin2005} or external forcing \cite{Bao2020}. Here we present the first experimental observation of oscillatory antiphase switching between counter-propagating light beams in a passive Kerr resonator -- whereby the two fields exchange dominance. In photonics, noisy \cite{Giacomelli99} and chaotic \cite{Virte12} switching between two polarization states as well as inphase and antiphase frequency combs \cite{Hillbrand20} have been described in semiconductor lasers. Recently, similar effects have also been described in the simulation of driven-dissipative dimers of Bose-Einstein condensates \cite{Giraldo2020}. 

In our passive device, antiphase oscillations follow symmetry breaking between counter-propagating beams where one mode becomes dominant with respect to the other. Attractor merging then restores the broken symmetry on average by chaotically or periodically switching between the dominant components. We numerically model this self-switching antiphase behaviour using a coupled dynamical system that features only the self- and cross-phase Kerr nonlinearities and provide an explanation for its occurrence. This demonstrates that, despite the inevitable presence of other nonlinear phenomena in the microresonator -- including thermal nonlinearities, dispersion and small amounts of frequency comb generation \cite{DelHaye2007} -- the periodic and chaotic switching between the two counter-propagating beams is induced by Kerr nonlinearities describing self- and cross-phase modulation. 
Controllable switching between the two beams finds natural applications in encoding of optical signals, chaotic cryptography and waveform synthetisers. In addition to passive systems, our results may be of consequence for the gain dynamics of ring lasers \cite{Gelens2009,Trita2013}, and for systems that host Kerr solitons \cite{Kippenberg2018, Bao2019, Guo2016, Jang2018} -- especially those with counter-propagating modes \cite{Yang2017}.

\begin{figure}[h!]
\fbox{\includegraphics[scale=0.6]{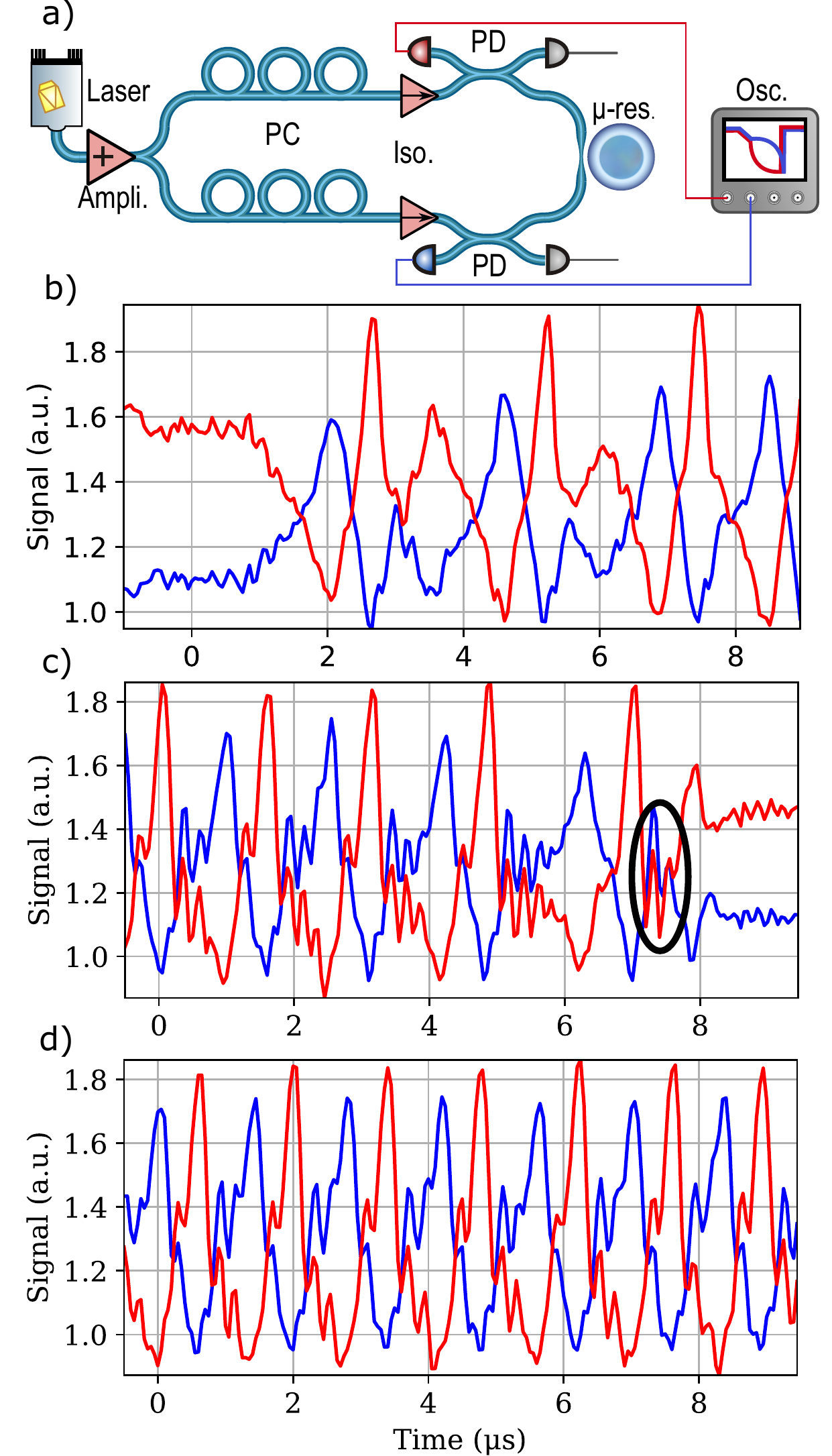}} 
\caption{a) Schematic of the experimental set-up. Ampli. = fibre amplifier, PC = polarization controllers, Iso. = optical isolators, PD = photodiodes, \textmu-res. = microresonator, Osc. = oscilloscope. Power attenuation components are omitted for simplicity. b)--d) Examples of different experimentally-observed self-switching behaviours under a detuning scan of a single cavity resonance -- the traces depict fluctuations in optical power. b) Chaotic intermittent switching where the modes do not exchange dominance every time. c) Transient synchronisation phenomena in-between transition events (circled). d) Weakly chaotic dynamics where the amplitudes of antiphase intensity oscillations reach similar minima and maxima between regular mode switches.}
\label{fig:exptTraces}
\end{figure}

The experimental set-up uses a microrod resonator (1.9 mm in diameter), with a cavity half-linewidth, $\gamma$, of approximately 1 MHz, machined from a rod of fused silica. The resonator is pumped at 1.55 $\mu$m with an amplified external-cavity diode laser, coupled via a tapered optical fibre. The optical circuitry used to pump and read the transmission spectra from this resonator, is depicted in Fig.~\ref{fig:exptTraces}a). The number of components is minimised in order to reduce optical losses, thereby maximising the power available to couple into the resonator, and so allowing easy access to symmetry-broken oscillatory and chaotic regimes. The power coupled into the resonator is inferred by subtracting the transmission signal in each direction from its respective baseline value -- measured with the laser out of resonance.

The input laser is scanned across the chosen cavity resonance. The polarization of each input branch is independently adjusted to maximise coupling efficiency and their relative powers are adjusted until the system begins spontaneously flipping between dominant directions of coupled light, indicating that they are sufficiently balanced to achieve spontaneous symmetry breaking. The oscillations within the symmetry-broken region are then investigated. Knowing that the input powers are balanced, the traces for each direction are subsequently re-scaled to share a common zero and maximum coupled power, in order to correct for differences in the responses of the photodiodes.
\begin{figure*}[htbp]
\centering
\fbox{\includegraphics[width=\textwidth]{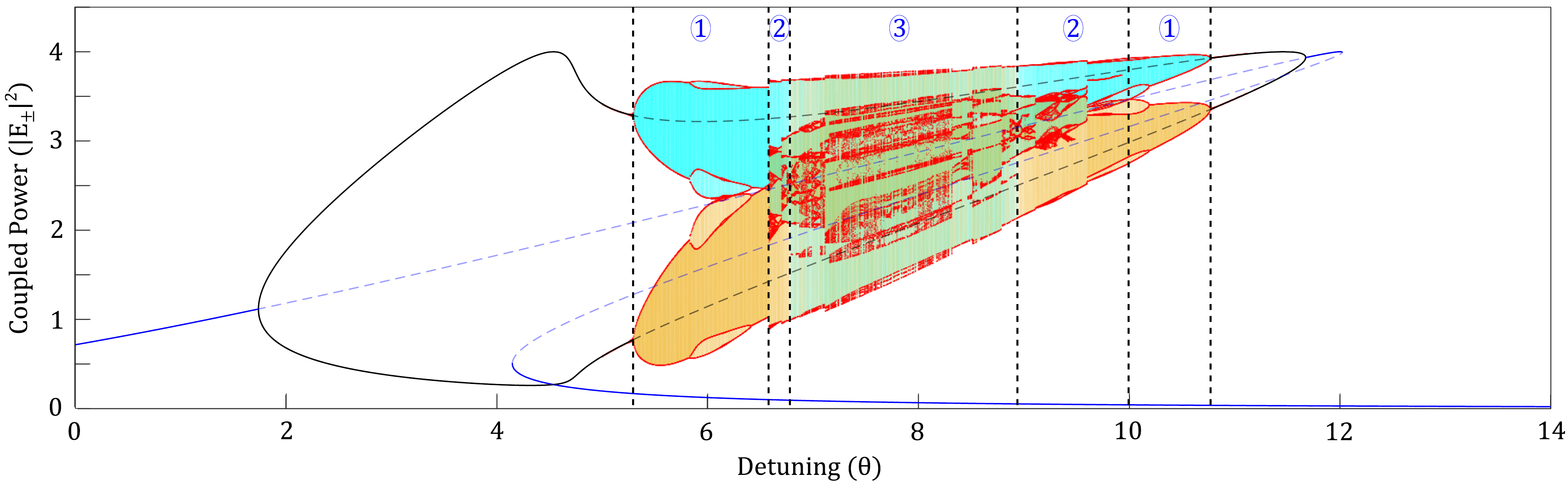}}
\caption{Poincar\'{e} sections from a forward scan of the detuning are overlaid atop the stationary solution set to Eq.~\eqref{eq:CoupEnvEvolution} with $|E_{\mathrm{in}}|^{2} = 4.0$. The blue and black lines indicate the symmetric and asymmetric stationary solutions, respectively, with solid lines indicating stable states and pale dashed lines indicating unstable states. The red dots denote the Poincar\'{e} sections - a red dot is placed at each local minimum and maximum of an oscillation. Also shown, in semitransparent yellow and blue, are the full ranges of the oscillations. The semitransparent yellow and blue shading results in a green colouring where the oscillations overlap. One observes three distinct regions: In zones (1), antiphase intensity oscillations do not overlap. In zones (2), they partially overlap, but this does not cause switching. In zone (3), two asymmetric attractors merge after symmetry restoring crises and antiphase switching occurs between the two counter-propagating intensities.}
\label{fig:DualPoinStat}
\end{figure*}
During the input laser scans several regimes of oscillatory antiphase switching are observed as displayed in Fig.~\ref{fig:exptTraces} b)-d) :

\textit{Chaotic switching of antiphase oscillations:} Chaotic switching is characterised by irregular exchanges of the dominant mode, with chaotic variations of the mode amplitudes (see Fig.~\ref{fig:exptTraces}~b)). 

\textit{Intermittent switching and transient synchronisation:} Another phenomenon that can occur is transient synchronisation of the mode switching followed temporarily by a loss of their antiphase relationship while spiralling towards the unstable symmetric solution, over a time-scale of roughly the inverse of the half-linewidth, $1/\gamma$. Since the symmetric state is unstable, the mode amplitudes subsequently fly apart. This is shown towards the right-hand side of Fig.~\ref{fig:exptTraces}c).

\textit{Weak chaos within periodic switching:} Weakly chaotic evolution between regular mode switching is observed in Fig.~\ref{fig:exptTraces}d). In these regions, fully periodic switching dynamics may be prevented by ineliminable internal noise.

The coupled equations used to model the counter-propagating field envelopes assume that they interact with each other simply via the Kerr nonlinearity, whereby the refractive index experienced by the two beams is modulated by the intensities of both. This system comprises two coupled complex differential equations:
\begin{equation}
\frac{d E_{\pm}}{d t}=E_{\mathrm{in}}-\left[1+i(\theta -A\left|E_{\pm}\right|^{2}-B\left|E_{\mp}\right|^2)\right]E_{\pm}\;,
\label{eq:CoupEnvEvolution}
\end{equation}
\noindent where $E_{\pm}$ are the field envelopes of the clockwise and counter-clockwise propagating fields seen in Fig.~\ref{fig:CounterFig}, $E_{\mathrm{in}}$ is the envelope of the input beam(s), $\theta$ is the cavity detuning (the difference between the frequency of the input beam and the closest cavity resonant frequency), and finally, $A$ and $B$ are the self- and cross-phase modulation constants, respectively. The system is symmetric upon exchange of $E_{\pm}$ with $E_{\mp}$ and the simplest symmetric solutions are those where the two components follow identical trajectories. Symmetry breaking in this system has been recently investigated in Refs.~\cite{woodley2018universal, Hill2020,Garbin19}. The relevant conversion from physical units for this model can be found in the SI to Ref.~\cite{del2017symmetry}.

Assuming the set-up of Fig.~\ref{fig:CounterFig}, we consider $A=1$ and $B=2$, since our resonator is non-diffusive \cite{kaplan1982directionally,firth1985diffusion}. Equations \eqref{eq:CoupEnvEvolution} are mathematically equivalent to those describing two co-propagating light components circulating within a Kerr ring resonator, with left and right circular polarizations, respectively \cite{geddes1994polarisation,woodley2018universal,Hill2020}. Although we experimentally focus on the case of counter-propagating fields, similar effects to those described here should be observable in other systems such as in Ref.~\cite{martin2010homogeneous}.

The susceptibility of this system to oscillations -- and the amplitude of such oscillations -- increases with respect to both the input power and the ratio of the strengths of cross-phase and self-phase modulation ($B/A$) although sudden crises between the oscillations and unstable stationary solutions can occur \cite{Hill2020}.
Equations \eqref{eq:CoupEnvEvolution} have been shown to exhibit chaotic behaviour after a rapid succession of period-doubling bifurcations \cite{woodley2018universal, Hill2020}. It is, therefore, instructive to visualise this system with Poincar\'{e} sections that sample the maxima and minima of the intensity oscillations  -- see Fig.~\ref{fig:DualPoinStat}.

When in the chaotic state, one can, by increasing the detuning, scan into and out of regimes of oscillatory antiphase switching between the two field components. Before and after the transitions to the self-switching region, there are two asymmetric attractors, mirror images of each other (upon the transformation $E_{\pm} \rightarrow E_{\mp}$) that partially overlap and occupy similar regions of phase space (zone (2) in Fig.~\ref{fig:DualPoinStat}). At the transitions, symmetry restoration crises \cite{Ben-Tal2002} take place and the two asymmetric attractors merge into a single one where the symmetry is restored on average. Inside zone (3) in Fig.~\ref{fig:DualPoinStat}, chaotic, intermittent and periodic self-switching take place on a single attractor that contains both asymptotic evolutions of $E_{\pm}$ and $E_{\mp}$, i.e. one remains on the same attractor upon the transformation $E_{\pm} \rightarrow E_{\mp}$. This implies that the originally broken symmetry is restored on average over the period of the main component of oscillation.

\begin{figure}[t!]
\centering
\fbox{\includegraphics[width=0.85\linewidth]{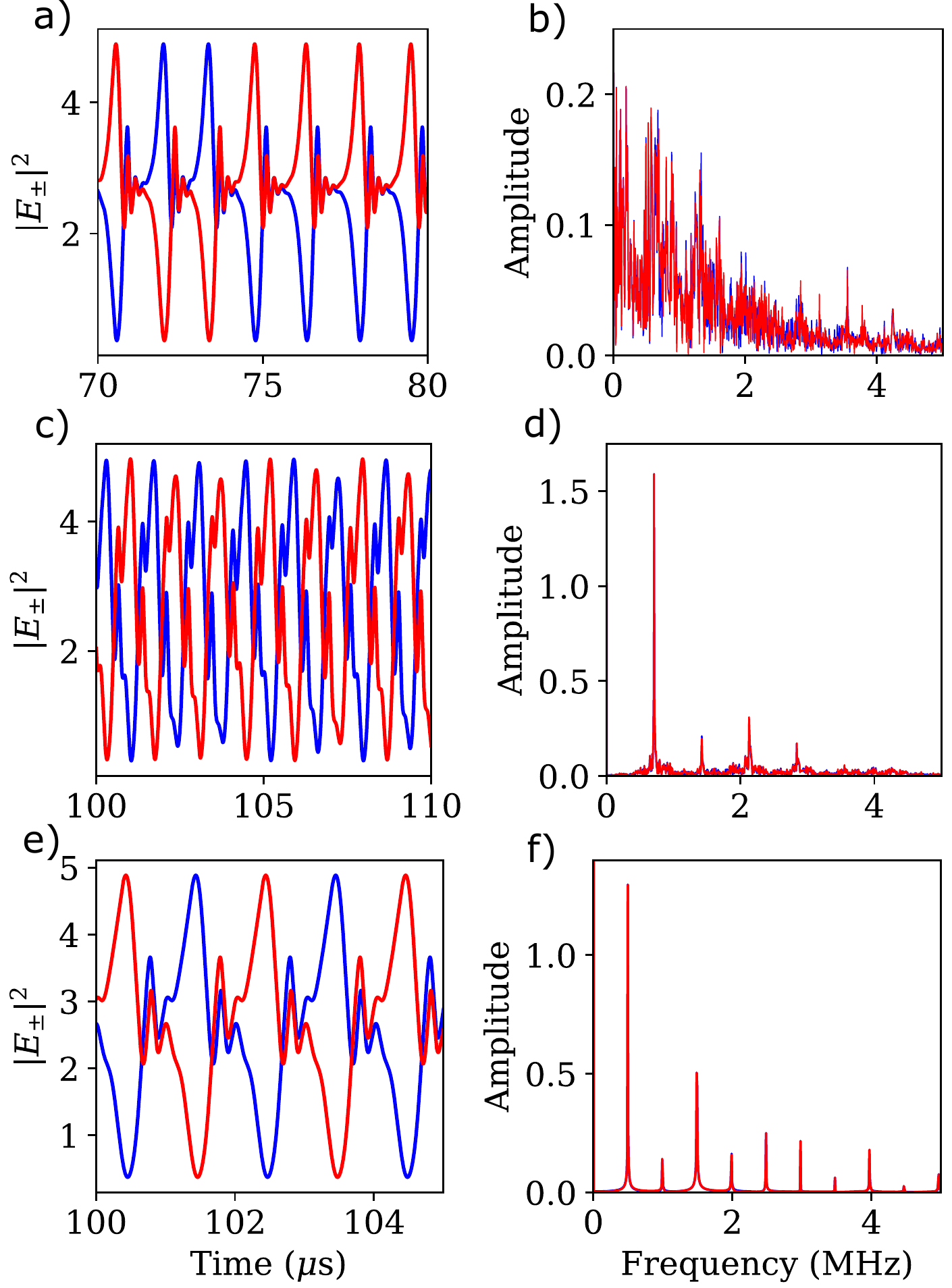}}
\caption{Time series and rf intensity spectra (FFT of $|E_{\pm}|^{2}$) of different switching states obtained by numerical integration of Eq.~\eqref{eq:CoupEnvEvolution} for $|E_{\mathrm{in}}|^{2} = 5.8$. a)-b) Chaotic switching: the modes swap irregularly, with chaotic amplitude variations for, $\theta = 6.984$. c)-d) Weak chaotic evolution within periodic switches for $\theta = 7.1$. e)-f) Periodic switching which implies disappearance of broadband chaos for $\theta = 7.004$. }
\label{fig:phasePlots}
\end{figure}

For appropriate values of detuning and input power, we numerically reproduce the essential features of the experimental traces that display self-switching as shown in Fig.~\ref{fig:exptTraces}. We first select $|E_{\mathrm{in}}|^{2} = 5.8$ to match the range of switching frequencies in the experiment corresponding to $\gamma$ around 1 MHz. Fig.~\ref{fig:phasePlots} a), c) and e) show examples of mode switching oscillations of the intensities while the corresponding spectra are displayed in Fig.~\ref{fig:phasePlots} b), d) and f). 
Intermittent switching in Fig.~\ref{fig:phasePlots} a) and b) results in broadband chaotic dynamics of the antiphase oscillations of the mode intensities. For $\theta = 7.1$ in Fig.~\ref{fig:phasePlots} c) and d) we observe regular switching between the two intensities although the amplitudes of the modes evolve in a weakly chaotic way (see the lower part of the power spectrum). Small adjustments of the detuning parameter lead to fully periodic self-switching between the counter-propagating modes (see Fig.~\ref{fig:phasePlots} e) and f)). This last regime is difficult to observe experimentally because of internal noise and residual asymmetries. 

\begin{figure}[t!]
\centering
\fbox{\includegraphics[width=0.85\linewidth]{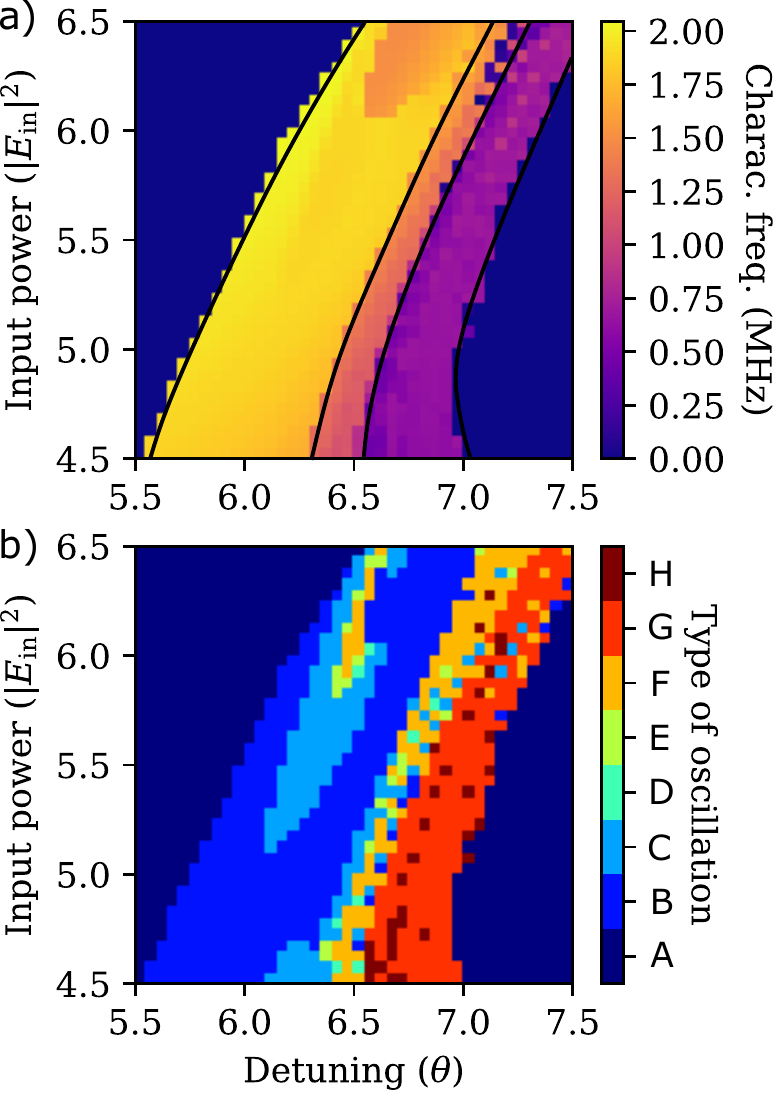}}
\caption{Heat-maps when varying the pump power and detuning. a) Characteristic frequencies (key changes in behaviour are marked with black curves). b) Various types of dynamics: A - steady state; B - regular oscillations; C - period-doubled; D - period-tripled; E - period-quadrupled; F - chaotic, non-switching; G - chaotic switching; H - periodic switching. The period-tripled state, like periodic switching, is associated with the total disappearance of chaos. The system reverts to steady state behaviour towards the right of the frames.}
\label{fig:heatMap}
\end{figure}

The robustness and wide parameter range of self-switching Kerr oscillations in microresonators is presented in Fig.~\ref{fig:heatMap} where we compare and contrast the different dynamical regimes when changing the detuning $\theta$ and the input power $|E_{\mathrm{in}}|^{2}$. In Fig.~\ref{fig:heatMap} a) we report the characteristic frequency by extracting the dominant Fourier component at non-zero frequency from rf intensity spectra such as those displayed in Fig.~\ref{fig:phasePlots} b), d) and f). In the region of mode switching, the characteristic frequency lowers -- see the purple area in Fig.~\ref{fig:heatMap} a), to around half of the typical values before the onset of switching, another signature of the different dynamical character of the switching and non-switching oscillations. In Fig.~\ref{fig:heatMap} b) where different dynamical behaviours are characterized by different colours, red regions of self-switching are clearly visible. Here, periodic oscillations such as those displayed in Fig.~\ref{fig:phasePlots} e) and f) are sporadic and occur through crises of chaotic attractors.

We have experimentally observed self-switching behaviour in counter-propagating light, including chaotic and almost periodic switching, using a passive microresonator with Kerr nonlinearity. We have also demonstrated that these behaviours can be explained by a simple dynamical system that considers only Kerr effects. In particular, the self-switching regime occurs in a well-defined region of the parameter space delimited by a symmetry restoring (on average) global bifurcation. This model also reproduces all the observed dynamical states of self-switching in the microrod resonator -- including excellent agreement with the observed frequency of switching -- as well as predicting the total disappearance of chaos in the form of periodic switching. Our results are of interest in the study of global bifurcations in dynamical systems such as symmetry-restoring crisis. From a practical perspective, self-switching periodicity of the counter-propagating modes can be applied in the controlled generation of twin waveforms and signal encoding while chaotic states can be potentially employed in the generation of chaotic-cryptographic algorithms \cite{Amigo2007} as well as chaos-induced stochastic resonance \cite{Monifi16}.

We acknowledge financial support from: Engineering and Physical Sciences Research Council (EPSRC) DTA Grant No. EP/M506643/1; H2020 Marie
Sk\l{}odowska-Curie Actions (MSCA) (748519, CoLiDR); National Physical Laboratory Strategic Research; H2020 European Research Council (ERC) (756966, CounterLight); EPSRC and the Scottish Universities Physics Alliance (SUPA). We would also like to thank Niall Moroney for his advice and assistance with some of the coding.

$\dagger$ M.T.M.W. and L.H. contributed equally to this work and should be considered as co-first authors.

\bibliography{Ref12}

\end{document}